\documentclass[preprint,aps,pra,showpacs,floatfix]{revtex4}
\usepackage{graphicx}
\usepackage{nicefrac}
\usepackage{amsmath}
\usepackage{amsfonts}
\usepackage{amssymb}
\usepackage{amsthm}
\usepackage{epsf}
\usepackage{bm}
\usepackage{dcolumn}
\newcolumntype{.}{D{x}{}{-1}}

\newcommand{\be}{\begin{eqnarray}}
\newcommand{\ee}{\end{eqnarray}}

\newcommand{\phm}{\phantom{$-$}}

\newcommand{\mun}{\mu_N}
\newcommand{\la}{\langle}
\newcommand{\ra}{\rangle}

\newcommand{\bfr}{{\bf r}}

\newcommand{\bfT}{{\bf T}}

\newcommand{\balpha}{ {\bm\alpha} }

\newcommand{\bmu}{\bm\mu}

\newcommand{\aZ}{\alpha Z}

\newcommand{\ket}[1]{|#1\rangle}

\newcommand{\matrixel}[3]{\langle #1 | #2 | #3 \rangle}
\newcommand{\epssph}{\varepsilon[\text{sph}]}
\newcommand{\epsws}{\varepsilon[\text{WS}]}
\newcommand{\xQED}{x_\text{QED}}
\newcommand{\xSE}{x_\text{SE}}
\newcommand{\xVP}{x_\text{VP}}
\begin{document}
\title{
Ground-state hyperfine splitting of B-like ions in high-$Z$ region}
\author{
D.~A.~Glazov,$^{1,2,3}$ A.~V.~Volotka,$^{2,3}$ O.~V.~Andreev,$^{1}$ V.~P.~Kosheleva,$^{1,2,3}$ S.~Fritzsche,$^{2,3,4}$ V.~M.~Shabaev,$^{1}$ G.~Plunien,$^{5}$ and Th.~St\"ohlker$^{2,3,6}$
}
\affiliation{
$^1$
Department of Physics, St. Petersburg State University, 199034 St. Petersburg, Russia \\
$^2$
Helmholtz-Institut Jena, D-07743 Jena, Germany \\
$^3$
GSI Helmholtzzentrum f\"ur Schwerionenforschung GmbH, D-64291 Darmstadt, Germany \\
$^4$
Theoretisch-Physikalisches Institut, Friedrich-Schiller-Universit\"at Jena, D-07743 Jena, Germany \\
$^5$
Institut f\"ur Theoretische Physik, Technische Universit\"at Dresden, D-01062 Dresden, Germany \\
$^6$
Institut f\"ur Optik und Quantenelektronik, Friedrich-Schiller-Universit\"at, D-07743 Jena, Germany
}
\begin{abstract}
The hyperfine splitting of the ground state of selected B-like ions within the range of nuclear charge numbers $Z=49$--$83$ is investigated in detail. The rigorous QED approach together with the large-scale configuration-interaction Dirac-Fock-Sturm method are employed for the evaluation of the interelectronic-interaction contributions of first and higher orders in $1/Z$. The screened QED corrections are evaluated to all orders in $\aZ$ by using an effective potential. The influence of nuclear magnetization distribution is taken into account within the single-particle nuclear model.
\end{abstract}


\maketitle

%
\section{Introduction}
%
%
Investigations of the hyperfine splitting (hfs) in highly charged ions provide an access to the bound-state QED effects in the presence of strong electric and magnetic nuclear field. Several experiments were performed with hydrogenlike ions \cite{klaft:94:prl,crespo:96:prl,crespo:98:pra,seelig:98:prl,beiersdorfer:01:pra} accompanied by the progress in theory \cite{schneider:94:pra,shabaev:94:jpb,shabaeva:95:pra,persson:96:prl,shabaev:97:pra,blundell:97:pra,shabaev:98:pra,sunnergren:98:pra}. However, the QED contribution appears to be obscured by the Bohr-Weisskopf (nuclear magnetization distribution) effect. In order to overcome this problem, it was proposed to consider a specific difference of the hfs values in hydrogenlike and lithiumlike ions with the same nucleus \cite{shabaev:01:prl}. Further theoretical investigations of the hfs in highly charged ions continued the last two decades \cite{boucard:00:epjd,shabaev:00:hi,artemyev:01:pra,sapirstein:01:pra,yerokhin:01:pra,sen'kov:02:npa,volotka:03:epjd,sapirstein:06:pra,oreshkina:07:os,kozhedub:07:pra,sapirstein:08:pra,volotka:08:pra,oreshkina:08:pla}. On top of that, the rigorous evaluation of the two-electron QED diagrams --- screened self-energy, screened vacuum-polarization and two-photon exchange --- provided the most accurate up-to-date value of the specific difference for Li-like and H-like bismuth $^{209}\mathrm{Bi}$ \cite{volotka:09:prl,glazov:10:pra,andreev:12:pra,volotka:12:prl}. Following a few less successful attempts, recent experimental efforts at GSI finally lead to the measurement of the hfs in Li-like bismuth \cite{lochmann:14:pra,ullmann:15:jpb,sanchez:17:jpb}. Surprisingly, the experimental and theoretical values have been found $7\sigma$ apart \cite{ullmann:17:nc} which established the so-called ``Hyperfine Puzzle'' \cite{karr:17:nphys}. The solution to this problem has been found in the strongly underestimated uncertainty of the tabulated value of the nuclear magnetic moment of $^{209}\mathrm{Bi}$. Recent calculation of the magnetic shielding constant relevant for the nuclear magnetic resonance method and the corresponding independent measurement have provided a new value of the nuclear magnetic moment and thus restored the agreement between theory and experiment \cite{skripnikov:18:prl}. However, the present agreement is based on the theoretical calculations of the magnetic shielding constant and in view of the previous situation can not serve alone as a sufficiently reliable test of the bound-state QED. To eliminate different possible explanations of the discrepancy, which can coexist, an independent investigation with another bismuth isotope $^{208}\mathrm{Bi}$ has been proposed \cite{schmidt:18:plb}.

Another way to reinforce the verification of the underlying theory by the experiment is to consider different charge states, e.g., boronlike ions, in addition to hydrogenlike and lithiumlike ones. Previously, the theoretical values of the hfs in boronlike ions have been reported in the middle-$Z$ region \cite{volotka:08:pra} and for two heavy ions, lead and bismuth \cite{oreshkina:08:pla}. The hyperfine constants for the ground and several excited states in the middle-$Z$ region have been evaluated within the MCDHF approach in Ref.~\cite{verdebout:14:adndt}. The QED corrections for the $2p_j$ states have been addressed in Refs.~\cite{sapirstein:06:pra,sapirstein:08:pra}. In the present paper, we perform comprehensive relativistic calculations of the interelectronic-interaction and QED contributions to the ground-state hfs for selected boronlike ions in the range $Z=49$--$83$. The rigorous QED perturbation theory in the first order in $1/Z$ is combined with the all-order CI-DFS calculations to account for the correlation effects. The one-loop radiative corrections are evaluated to all orders in $\aZ$ with an effective local screening potential. The Bohr-Weisskopf effect is investigated within the single-particle nuclear model. As a result, the most accurate up-to-date hfs values are obtained. Thus we eliminate the present deficiency of the theoretical data for heavy boronlike ions.
%

The relativistic units ($\hbar=1$, $c=1$) and the Heaviside charge unit \nobreak[${\alpha=e^2/(4\pi),e<0}$] are used throughout the paper.

%
\section{Theoretical methods}
\label{sec:theory}
%
%
The interaction of atomic electrons with the nuclear magnetic moment is described by the Fermi-Breit operator,
\begin{equation}
  H_\mu = \frac{|e|}{4\pi} \bmu \cdot \bfT
\,,
\end{equation}
where $\bmu$ is the nuclear magnetic moment operator. The electronic operator $\bfT$ is given by
\begin{equation}
\label{eq:T}
  \bfT = \sum_i \frac{[\bfr_i \times \balpha_i]}{r_i^3} F(r_i)
\,,
\end{equation}
where the summation runs over the atomic electrons, $\balpha$ is the Dirac-matrix vector, and $F(r)$ is the nuclear magnetization distribution factor discussed below. The ground-state hfs of a highly charged boronlike ion can be written as
\begin{align}
\label{eq:E_hfs}
  \Delta E & = \alpha (\aZ)^3 g_I \, \frac{m_e}{m_p} \, \frac{2I+1}{36} \, \frac{1}{(1+m_e/M)^3}
\nonumber\\
    & \times \left[ A(\aZ)(1-\delta)(1-\varepsilon) + \frac{1}{Z}B(\aZ) + \frac{1}{Z^2}C(Z,\aZ) + \xQED \right]
\,.
\end{align}
Here $g_I=\mu/(\mu_N I)$ is the $g$ factor of the nucleus with magnetic moment $\mu$ and spin $I$, $\mu_N$ is the nuclear magneton, and $m_e$, $m_p$, and $M$ are the electron, proton and nuclear masses, respectively. The one-electron point-nucleus relativistic factor $A(\alpha Z)$ is modified by the finite nuclear charge distribution ($\delta$) and by the finite nuclear magnetic moment distribution ($\varepsilon$). The interelectronic-interaction correction is represented by the first-order term $B(\aZ)/Z$ and by the higher-order term $C(Z,\aZ)/Z^2$. The QED corrections are represented by $\xQED$. These contributions to the hfs are individually considered below.

%
%
\subsection{One-electron contributions}
\label{sec:one-el}
%
%
When the nuclear parameters and the nonrelativistic value of the hfs is separated out it can be found that the one-electron contributions can be numerically evaluated in terms of a matrix element of the zero component $T_0$ of the operator $\bfT$ given by Eq.~(\ref{eq:T}),
\begin{equation}
  A(\aZ)(1-\delta)(1-\varepsilon) = \frac{9}{(\aZ)^3 M_j} \, \matrixel{a}{T_0}{a}
\,.
\end{equation}
The wave function $\ket{a}$ of the valence $2p_{1/2}$ state with the total angular momentum projection $M_j$ is obtained as a solution of the Dirac equation with the potential of the extended nucleus.

The relativistic factor $A(\alpha Z)$ for the point-like nucleus is known analytically,
\begin{equation}
  A(\alpha Z) = \frac{24}{N^3 (N+1) \gamma (2 \gamma -1)}
\,,
\end{equation}
where $\gamma = \sqrt{1-(\aZ)^2}$ and $N = \sqrt{2\gamma+2}$.
The nuclear charge distribution correction $\delta$ can be found either analytically \cite{shabaev:94:jpb,volotka:03:epjd} or numerically by solving the Dirac equation with the extended nucleus potential. In the present work $\delta$ is found numerically using the Fermi model for the nuclear charge distribution. In order to estimate the uncertainty due to the model dependence of the nuclear charge distribution we calculate $\delta$ also for the homogeneously charged sphere model. The Bohr-Weisskopf correction $\varepsilon$ is treated by introducing some volume distribution function $F(r)$ instead of the point-nucleus value $F(r)=1$. For the homogeneous sphere model
\begin{equation}
\label{eq:F_sph}
  F(r) =
    \begin{cases}
      (r/R_0)^3, &  r \leq R_0 \,,\\
      1, &  r > R_0 \,,
    \end{cases}
\end{equation}
where $R_0=\sqrt{5/3}{\langle r^2 \rangle}^{1/2}$ is the radius of the sphere related to the root-mean-square charge radius. Within the more sophisticated single-particle model, the nuclear magnetic moment is determined by the total angular momentum of the unpaired nucleon (proton or neutron) \cite{bellac:63:np,shabaev:97:pra}. Then the nuclear $g_I$ factor is just the Land\'e factor of an extra nucleon, which is defined by the well-known formula
\begin{equation}
\label{eq:g_I}
  g_I = \mu/\mu_N I = \frac{1}{2}\left[(g_L+g_S) + (g_L-g_S) \frac{L(L + 1)-3/4}{I(I + 1)}\right]
\,,
\end{equation}
where $L$ is the nuclear orbital momentum, $g_L$ and $g_S$ are the orbital and spin $g$ factors of the valence nucleon, respectively. This equation is employed to find $g_S$ such as to reproduce the experimental value of the nuclear magnetic moment $\mu=\mu_N I g_I$, while $g_L$ is chosen as $g_L = 1$ for the unpaired proton and $g_L = 0$ for the unpaired neutron. The odd-nucleon wave function is found as a solution of the Schr\"odinger equation with the Woods-Saxon potential~\cite{woods:54:pr,rost:68:plb} and with account for the spin-orbit interaction. All the formulas needed for these calculations can be found, e.g., in Ref.~\cite{shabaev:97:pra}. Following Ref.~\cite{shabaev:97:pra} we assign the uncertainty of $\varepsilon$ found within single-particle model as follows. For the ions where the spin and orbital parts of the Bohr-Weisskopf correction are of the same sign the uncertainty is estimated to be about 30\% of $\varepsilon$. For the ions where the spin and orbital parts are of opposite sign the uncertainty is defined as 20\% of the Bohr-Weisskopf correction evaluated within the homogeneous sphere model. The ions of lead and bismuth are treated in a special way: the $1s$ results for $^{207}$Pb and $^{209}\mathrm{Bi}$ from Ref.~\cite{sen'kov:02:npa} are rescaled to obtain the Bohr-Weisskopf correction for the $2p_{1/2}$ state (see also Refs.~\cite{shabaev:01:prl,oreshkina:08:pla}).
%
%
\subsection{Many-electron contributions}
\label{sec:many-el}
%
%
The interelectronic interaction can be taken into account within the Breit approximation by averaging the operator $T_0$ with the many-electron wave function evaluated within one of the available methods. We use the large scale configuration interaction method in the basis of the Dirac-Fock-Sturm orbitals (CI-DFS) \cite{bratsev:77} which has been intensively developed and successfully employed, in particular, for the hfs calculations in few-electron ions \cite{oreshkina:07:os,volotka:08:pra,oreshkina:08:pla,volotka:12:prl}. Since the configuration interaction is implemented in the space of the positive-energy Dirac states, the contribution of the negative-energy Dirac states needs to be taken into account separately \cite{tupitsyn:05:pra}. The results obtained within the CI-DFS method can be improved by the rigorous evaluation of the leading orders of the bound-state QED perturbation theory. In this work, we evaluate the first-order contribution represented by the one-photon-exchange diagrams. The corresponding formulas were derived in Ref.~\cite{shabaeva:95:pra}. The numerical calculations are performed using the dual-kinetic-balance method \cite{shabaev:04:prl} with basis functions constructed from B splines \cite{sapirstein:96:jpb}. This yields the first-order term $B(\aZ)/Z$ in Eq.~(\ref{eq:E_hfs}). The corresponding term within the Breit approximation is subtracted from the CI-DFS result to obtain the contribution of the second and higher orders $C(Z,\aZ)/Z^2$. The effects beyond the Breit approximation are estimated to be of the order $(\aZ)^3 C(Z,\aZ)/Z^2$. The finite distributions of the nuclear charge and magnetic moment are taken into account using the extended nucleus potential and the nuclear magnetization distribution factor $F(r)$, in the same way as for the one-electron contributions.
%
%
\subsection{QED contributions}
\label{sec:QED}
%
%
The radiative correction $\xQED$ of the first order in $\alpha$ is the sum of the self-energy (SE) and vacuum-polarization (VP) contributions, $\xQED = \xSE + \xVP$. Rigorous evaluation of these terms for the $p_{1/2}$ states was accomplished in Ref.~\cite{sapirstein:06:pra} with the Coulomb potential and with the Kohn-Sham local screening potential for neutral atoms. High numerical precision was achieved for low-$Z$ hydrogenlike ions in Ref.~\cite{yerokhin:10:pra}. In Ref.~\cite{volotka:08:pra} this correction was calculated for the $2p_{1/2}$ state with the Kohn-Sham potential for boronlike ions in the range $Z=7-28$ and in Ref.~\cite{oreshkina:08:pla} for several high-$Z$ ions. We extend these calculations to boronlike ions in the range $Z=49-83$. In order to account approximately for the many-electron QED contributions we introduce an effective screening potential in the Dirac equation. The well-known core-Hartree (CH) and Kohn-Sham (KS) potentials are used, see, e.g., our previous studies \cite{volotka:08:pra,oreshkina:08:pla}. For the final value we take the KS results, while the uncertainty is estimated as 50\% of the screening effect (difference between the KS and Coulomb values). This rather conservative estimation is supposed to cover the significant contribution of the two-electron diagrams with the SE or VP loop connected to the $1s$ or $2s$ electron. This contribution is supposed to be relatively larger than, e.g., in lithiumlike ions. In contrast to the diagrams with the SE or VP loop connected to the $2p$ electron, it is not taken into account by the screening potentials.
%
%
\section{Results and discussion}
\label{sec:results}
%
%
The obtained results for the one-electron contributions are presented in Table~\ref{tab:A}. The nuclear spins $I$, parities $\pi$, and magnetic moments $\mu/\mun$ are taken from Ref.~\cite{stone:05:adndt}. For the magnetic moments of holmium and bismuth we use $\mu/\mun$[$^{165}$Ho] = 4.177(5) \cite{gustavsson:98:pra} and $\mu/\mun$[$^{209}$Bi] = 4.092(2) \cite{skripnikov:18:prl}, respectively. It should be noted here that the uncertainties of the nuclear magnetic moments for other elements might be also larger than indicated in Ref.~\cite{stone:05:adndt} due to the chemical shift \cite{skripnikov:18:prl,gustavsson:98:pra}. The nuclear radii are taken from Ref.~\cite{angeli:13:adndt}. The finite-nuclear-size correction $\delta$ is calculated with the Fermi model for the nuclear charge distribution. For the Bohr-Weisskopf correction $\epsilon$, we present the results obtained from the simple homogeneous sphere model ($\epssph$) as well as from the single-particle model where the wave function is found with the use of the Woods-Saxon potential ($\epsws$). The latter is taken as the final value except for lead and bismuth where the results of Ref.~\cite{sen'kov:02:npa} are scaled to the $2p_{1/2}$ state. The uncertainty is estimated as described in Sec.~\ref{sec:one-el}. 

The one-loop QED correction $\xQED$ is conveniently expressed in terms of the function $D(\aZ)$ as
\begin{equation}
\label{eq:xD}
    \xQED = \frac{\alpha}{\pi}\,D(\aZ)
\,.
\end{equation}
Table~\ref{tab:QED} displays separately the results for the self-energy and vacuum-polarization contributions. These calculations have been performed with the Coulomb and two different screening potentials, CH and KS. For the final value of $\xQED$ the KS results are taken with the uncertainty assessed as described in Sec.~\ref{sec:QED}.

In Table~\ref{tab:total} the one-electron part is presented by the value of the first term $A'=A(\aZ)(1-\delta)(1-\varepsilon)$ in the brackets in Eq.~(\ref{eq:E_hfs}) together with its uncertainty which mainly originates from the Bohr-Weisskopf correction. The interelectronic-interaction contributions $B(\alpha Z)/Z$ and $C(Z,\alpha Z)/Z^2$ and the QED correction $\xQED$ are shown in columns 3--5. The total hfs value $\Delta E$ is calculated according to Eq.~(\ref{eq:E_hfs}) with the nuclear magnetic moment given in Table~\ref{tab:A}. In addition, we present the ``reduced hyperfine splitting'' $\Delta E/(\mu/\mun)$ in the last column of Table~\ref{tab:total}. The previously reported values for lead and bismuth ions from Ref.~\cite{oreshkina:08:pla} are also given for comparison. The main novelties of the present results are the different treatment of the magnetization distribution effect and the different uncertainty estimation for the $C(Z,\aZ)/Z^2$ term. As one can see from the Table, the uncertainty of final values almost equally originates from the Bohr-Weisskopf correction and from the higher-order interelectronic-interaction contribution. This, in particular, strongly motivates the rigorous evaluation of the two-photon exchange diagrams, which has been performed so far only for lithiumlike ions \cite{volotka:12:prl}.
%
%
\section{Conclusion}
\label{sec:conclusion}
%
%
The hyperfine splitting in heavy boronlike ions is evaluated within the fully relativistic approach. The interelectronic-interaction effects are taken into account to all orders in $1/Z$ within the Breit approximation, while the first-order term is evaluated within the rigorous QED approach. The one-loop QED corrections are calculated to all orders in $\aZ$ with the screening effect taken into account by the effective local potential. The Bohr-Weisskopf effect is considered within the single-particle nuclear model. As the result, the most accurate up-to-date values for the ground-state hyperfine splitting in boronlike ions in the range $Z=49-83$ are presented.
%
%
%
%
\acknowledgments
The work was supported in part by DFG (Grant No.~VO 1707/1-3), by SPbSU-DFG (Grant No.~11.65.41.2017 and No.~STO 346/5-1), and by RFBR (Grants No.~16-02-00334 and 19-02-00974).
%
%
%
%

\begin{thebibliography}{50}
\expandafter\ifx\csname natexlab\endcsname\relax\def\natexlab#1{#1}\fi
\expandafter\ifx\csname bibnamefont\endcsname\relax
  \def\bibnamefont#1{#1}\fi
\expandafter\ifx\csname bibfnamefont\endcsname\relax
  \def\bibfnamefont#1{#1}\fi
\expandafter\ifx\csname citenamefont\endcsname\relax
  \def\citenamefont#1{#1}\fi
\expandafter\ifx\csname url\endcsname\relax
  \def\url#1{\texttt{#1}}\fi
\expandafter\ifx\csname urlprefix\endcsname\relax\def\urlprefix{URL }\fi
\providecommand{\bibinfo}[2]{#2}
\providecommand{\eprint}[2][]{\url{#2}}

\bibitem[{\citenamefont{Klaft et~al.}(1994)\citenamefont{Klaft, Borneis, Engel,
  Fricke, Grieser, Huber, K{\"u}hl, Marx, Neumann, Schr{\"o}der
  et~al.}}]{klaft:94:prl}
\bibinfo{author}{\bibfnamefont{I.}~\bibnamefont{Klaft}},
  \bibinfo{author}{\bibfnamefont{S.}~\bibnamefont{Borneis}},
  \bibinfo{author}{\bibfnamefont{T.}~\bibnamefont{Engel}},
  \bibinfo{author}{\bibfnamefont{B.}~\bibnamefont{Fricke}},
  \bibinfo{author}{\bibfnamefont{R.}~\bibnamefont{Grieser}},
  \bibinfo{author}{\bibfnamefont{G.}~\bibnamefont{Huber}},
  \bibinfo{author}{\bibfnamefont{T.}~\bibnamefont{K{\"u}hl}},
  \bibinfo{author}{\bibfnamefont{D.}~\bibnamefont{Marx}},
  \bibinfo{author}{\bibfnamefont{R.}~\bibnamefont{Neumann}},
  \bibinfo{author}{\bibfnamefont{S.}~\bibnamefont{Schr{\"o}der}},
  \bibnamefont{et~al.}, \bibinfo{journal}{Phys. Rev. Lett.}
  \textbf{\bibinfo{volume}{73}}, \bibinfo{pages}{2425} (\bibinfo{year}{1994}).

\bibitem[{\citenamefont{{Crespo~L\'opez-Urrutia}
  et~al.}(1996)\citenamefont{{Crespo~L\'opez-Urrutia}, Beiersdorfer, Savin, and
  Widmann}}]{crespo:96:prl}
\bibinfo{author}{\bibfnamefont{J.~R.} \bibnamefont{{Crespo~L\'opez-Urrutia}}},
  \bibinfo{author}{\bibfnamefont{P.}~\bibnamefont{Beiersdorfer}},
  \bibinfo{author}{\bibfnamefont{D.~W.} \bibnamefont{Savin}}, \bibnamefont{and}
  \bibinfo{author}{\bibfnamefont{K.}~\bibnamefont{Widmann}},
  \bibinfo{journal}{Phys. Rev. Lett.} \textbf{\bibinfo{volume}{77}},
  \bibinfo{pages}{826} (\bibinfo{year}{1996}).

\bibitem[{\citenamefont{{Crespo~L\'opez-Urrutia}
  et~al.}(1998)\citenamefont{{Crespo~L\'opez-Urrutia}, Beiersdorfer, Widmann,
  Birkett, {M{\aa}rtensson-Pendrill}, and Gustavsson}}]{crespo:98:pra}
\bibinfo{author}{\bibfnamefont{J.~R.} \bibnamefont{{Crespo~L\'opez-Urrutia}}},
  \bibinfo{author}{\bibfnamefont{P.}~\bibnamefont{Beiersdorfer}},
  \bibinfo{author}{\bibfnamefont{K.}~\bibnamefont{Widmann}},
  \bibinfo{author}{\bibfnamefont{B.~B.} \bibnamefont{Birkett}},
  \bibinfo{author}{\bibfnamefont{A.-M.}
  \bibnamefont{{M{\aa}rtensson-Pendrill}}}, \bibnamefont{and}
  \bibinfo{author}{\bibfnamefont{M.~G.~H.} \bibnamefont{Gustavsson}},
  \bibinfo{journal}{Phys. Rev. A} \textbf{\bibinfo{volume}{57}},
  \bibinfo{pages}{879} (\bibinfo{year}{1998}).

\bibitem[{\citenamefont{Seelig et~al.}(1998)\citenamefont{Seelig, Borneis, Dax,
  Engel, Faber, Gerlach, Holbrow, Huber, K{\"u}hl, Marx
  et~al.}}]{seelig:98:prl}
\bibinfo{author}{\bibfnamefont{P.}~\bibnamefont{Seelig}},
  \bibinfo{author}{\bibfnamefont{S.}~\bibnamefont{Borneis}},
  \bibinfo{author}{\bibfnamefont{A.}~\bibnamefont{Dax}},
  \bibinfo{author}{\bibfnamefont{T.}~\bibnamefont{Engel}},
  \bibinfo{author}{\bibfnamefont{S.}~\bibnamefont{Faber}},
  \bibinfo{author}{\bibfnamefont{M.}~\bibnamefont{Gerlach}},
  \bibinfo{author}{\bibfnamefont{C.}~\bibnamefont{Holbrow}},
  \bibinfo{author}{\bibfnamefont{G.}~\bibnamefont{Huber}},
  \bibinfo{author}{\bibfnamefont{T.}~\bibnamefont{K{\"u}hl}},
  \bibinfo{author}{\bibfnamefont{D.}~\bibnamefont{Marx}}, \bibnamefont{et~al.},
  \bibinfo{journal}{Phys. Rev. Lett.} \textbf{\bibinfo{volume}{81}},
  \bibinfo{pages}{4824} (\bibinfo{year}{1998}).

\bibitem[{\citenamefont{Beiersdorfer et~al.}(2001)\citenamefont{Beiersdorfer,
  Utter, Wong, {Crespo~L\'opez-Urrutia}, Britten, Chen, Harris, Thoe, Thorn,
  Tr{\"a}bert et~al.}}]{beiersdorfer:01:pra}
\bibinfo{author}{\bibfnamefont{P.}~\bibnamefont{Beiersdorfer}},
  \bibinfo{author}{\bibfnamefont{S.~B.} \bibnamefont{Utter}},
  \bibinfo{author}{\bibfnamefont{K.~L.} \bibnamefont{Wong}},
  \bibinfo{author}{\bibfnamefont{J.~R.}
  \bibnamefont{{Crespo~L\'opez-Urrutia}}},
  \bibinfo{author}{\bibfnamefont{J.~A.} \bibnamefont{Britten}},
  \bibinfo{author}{\bibfnamefont{H.}~\bibnamefont{Chen}},
  \bibinfo{author}{\bibfnamefont{C.~L.} \bibnamefont{Harris}},
  \bibinfo{author}{\bibfnamefont{R.~S.} \bibnamefont{Thoe}},
  \bibinfo{author}{\bibfnamefont{D.~B.} \bibnamefont{Thorn}},
  \bibinfo{author}{\bibfnamefont{E.}~\bibnamefont{Tr{\"a}bert}},
  \bibnamefont{et~al.}, \bibinfo{journal}{Phys. Rev. A}
  \textbf{\bibinfo{volume}{64}}, \bibinfo{pages}{032506}
  (\bibinfo{year}{2001}).

\bibitem[{\citenamefont{Schneider et~al.}(1994)\citenamefont{Schneider,
  Greiner, and Soff}}]{schneider:94:pra}
\bibinfo{author}{\bibfnamefont{S.~M.} \bibnamefont{Schneider}},
  \bibinfo{author}{\bibfnamefont{W.}~\bibnamefont{Greiner}}, \bibnamefont{and}
  \bibinfo{author}{\bibfnamefont{G.}~\bibnamefont{Soff}},
  \bibinfo{journal}{Phys. Rev. A} \textbf{\bibinfo{volume}{50}},
  \bibinfo{pages}{118} (\bibinfo{year}{1994}).

\bibitem[{\citenamefont{Shabaev}(1994)}]{shabaev:94:jpb}
\bibinfo{author}{\bibfnamefont{V.~M.} \bibnamefont{Shabaev}},
  \bibinfo{journal}{J. Phys. B} \textbf{\bibinfo{volume}{27}},
  \bibinfo{pages}{5825} (\bibinfo{year}{1994}).

\bibitem[{\citenamefont{Shabaeva and Shabaev}(1995)}]{shabaeva:95:pra}
\bibinfo{author}{\bibfnamefont{M.~B.} \bibnamefont{Shabaeva}} \bibnamefont{and}
  \bibinfo{author}{\bibfnamefont{V.~M.} \bibnamefont{Shabaev}},
  \bibinfo{journal}{Phys. Rev. A} \textbf{\bibinfo{volume}{52}},
  \bibinfo{pages}{2811} (\bibinfo{year}{1995}).

\bibitem[{\citenamefont{Persson et~al.}(1996)\citenamefont{Persson, Schneider,
  Greiner, Soff, and Lindgren}}]{persson:96:prl}
\bibinfo{author}{\bibfnamefont{H.}~\bibnamefont{Persson}},
  \bibinfo{author}{\bibfnamefont{S.~M.} \bibnamefont{Schneider}},
  \bibinfo{author}{\bibfnamefont{W.}~\bibnamefont{Greiner}},
  \bibinfo{author}{\bibfnamefont{G.}~\bibnamefont{Soff}}, \bibnamefont{and}
  \bibinfo{author}{\bibfnamefont{I.}~\bibnamefont{Lindgren}},
  \bibinfo{journal}{Phys. Rev. Lett.} \textbf{\bibinfo{volume}{76}},
  \bibinfo{pages}{1433} (\bibinfo{year}{1996}).

\bibitem[{\citenamefont{Shabaev et~al.}(1997)\citenamefont{Shabaev, Tomaselli,
  K{\"u}hl, Artemyev, and Yerokhin}}]{shabaev:97:pra}
\bibinfo{author}{\bibfnamefont{V.~M.} \bibnamefont{Shabaev}},
  \bibinfo{author}{\bibfnamefont{M.}~\bibnamefont{Tomaselli}},
  \bibinfo{author}{\bibfnamefont{T.}~\bibnamefont{K{\"u}hl}},
  \bibinfo{author}{\bibfnamefont{A.~N.} \bibnamefont{Artemyev}},
  \bibnamefont{and} \bibinfo{author}{\bibfnamefont{V.~A.}
  \bibnamefont{Yerokhin}}, \bibinfo{journal}{Phys. Rev. A}
  \textbf{\bibinfo{volume}{56}}, \bibinfo{pages}{252} (\bibinfo{year}{1997}).

\bibitem[{\citenamefont{Blundell et~al.}(1997)\citenamefont{Blundell, Cheng,
  and Sapirstein}}]{blundell:97:pra}
\bibinfo{author}{\bibfnamefont{S.~A.} \bibnamefont{Blundell}},
  \bibinfo{author}{\bibfnamefont{K.~T.} \bibnamefont{Cheng}}, \bibnamefont{and}
  \bibinfo{author}{\bibfnamefont{J.}~\bibnamefont{Sapirstein}},
  \bibinfo{journal}{Phys. Rev. A} \textbf{\bibinfo{volume}{55}},
  \bibinfo{pages}{1857} (\bibinfo{year}{1997}).

\bibitem[{\citenamefont{Shabaev et~al.}(1998)\citenamefont{Shabaev, Shabaeva,
  Tupitsyn, Yerokhin, Artemyev, K{\"u}hl, Tomaselli, and
  Zherebtsov}}]{shabaev:98:pra}
\bibinfo{author}{\bibfnamefont{V.~M.} \bibnamefont{Shabaev}},
  \bibinfo{author}{\bibfnamefont{M.~B.} \bibnamefont{Shabaeva}},
  \bibinfo{author}{\bibfnamefont{I.~I.} \bibnamefont{Tupitsyn}},
  \bibinfo{author}{\bibfnamefont{V.~A.} \bibnamefont{Yerokhin}},
  \bibinfo{author}{\bibfnamefont{A.~N.} \bibnamefont{Artemyev}},
  \bibinfo{author}{\bibfnamefont{T.}~\bibnamefont{K{\"u}hl}},
  \bibinfo{author}{\bibfnamefont{M.}~\bibnamefont{Tomaselli}},
  \bibnamefont{and} \bibinfo{author}{\bibfnamefont{O.~M.}
  \bibnamefont{Zherebtsov}}, \bibinfo{journal}{Phys. Rev. A}
  \textbf{\bibinfo{volume}{57}}, \bibinfo{pages}{149} (\bibinfo{year}{1998}).

\bibitem[{\citenamefont{Sunnergren et~al.}(1998)\citenamefont{Sunnergren,
  Persson, Salomonson, Schneider, Lindgren, and Soff}}]{sunnergren:98:pra}
\bibinfo{author}{\bibfnamefont{P.}~\bibnamefont{Sunnergren}},
  \bibinfo{author}{\bibfnamefont{H.}~\bibnamefont{Persson}},
  \bibinfo{author}{\bibfnamefont{S.}~\bibnamefont{Salomonson}},
  \bibinfo{author}{\bibfnamefont{S.~M.} \bibnamefont{Schneider}},
  \bibinfo{author}{\bibfnamefont{I.}~\bibnamefont{Lindgren}}, \bibnamefont{and}
  \bibinfo{author}{\bibfnamefont{G.}~\bibnamefont{Soff}},
  \bibinfo{journal}{Phys. Rev. A} \textbf{\bibinfo{volume}{58}},
  \bibinfo{pages}{1055} (\bibinfo{year}{1998}).

\bibitem[{\citenamefont{Shabaev et~al.}(2001)\citenamefont{Shabaev, Artemyev,
  Yerokhin, Zherebtsov, and Soff}}]{shabaev:01:prl}
\bibinfo{author}{\bibfnamefont{V.~M.} \bibnamefont{Shabaev}},
  \bibinfo{author}{\bibfnamefont{A.~N.} \bibnamefont{Artemyev}},
  \bibinfo{author}{\bibfnamefont{V.~A.} \bibnamefont{Yerokhin}},
  \bibinfo{author}{\bibfnamefont{O.~M.} \bibnamefont{Zherebtsov}},
  \bibnamefont{and} \bibinfo{author}{\bibfnamefont{G.}~\bibnamefont{Soff}},
  \bibinfo{journal}{Phys. Rev. Lett.} \textbf{\bibinfo{volume}{86}},
  \bibinfo{pages}{3959} (\bibinfo{year}{2001}).

\bibitem[{\citenamefont{Boucard and Indelicato}(2000)}]{boucard:00:epjd}
\bibinfo{author}{\bibfnamefont{S.}~\bibnamefont{Boucard}} \bibnamefont{and}
  \bibinfo{author}{\bibfnamefont{P.}~\bibnamefont{Indelicato}},
  \bibinfo{journal}{Eur. Phys. J. D} \textbf{\bibinfo{volume}{8}},
  \bibinfo{pages}{59} (\bibinfo{year}{2000}).

\bibitem[{\citenamefont{Shabaev et~al.}(2000)\citenamefont{Shabaev, Artemyev,
  Zherebtsov, Yerokhin, Plunien, and Soff}}]{shabaev:00:hi}
\bibinfo{author}{\bibfnamefont{V.~M.} \bibnamefont{Shabaev}},
  \bibinfo{author}{\bibfnamefont{A.~N.} \bibnamefont{Artemyev}},
  \bibinfo{author}{\bibfnamefont{O.~M.} \bibnamefont{Zherebtsov}},
  \bibinfo{author}{\bibfnamefont{V.~A.} \bibnamefont{Yerokhin}},
  \bibinfo{author}{\bibfnamefont{G.}~\bibnamefont{Plunien}}, \bibnamefont{and}
  \bibinfo{author}{\bibfnamefont{G.}~\bibnamefont{Soff}},
  \bibinfo{journal}{Hyperfine Interact.} \textbf{\bibinfo{volume}{127}},
  \bibinfo{pages}{279} (\bibinfo{year}{2000}).

\bibitem[{\citenamefont{Artemyev et~al.}(2001)\citenamefont{Artemyev, Shabaev,
  Plunien, Soff, and Yerokhin}}]{artemyev:01:pra}
\bibinfo{author}{\bibfnamefont{A.~N.} \bibnamefont{Artemyev}},
  \bibinfo{author}{\bibfnamefont{V.~M.} \bibnamefont{Shabaev}},
  \bibinfo{author}{\bibfnamefont{G.}~\bibnamefont{Plunien}},
  \bibinfo{author}{\bibfnamefont{G.}~\bibnamefont{Soff}}, \bibnamefont{and}
  \bibinfo{author}{\bibfnamefont{V.~A.} \bibnamefont{Yerokhin}},
  \bibinfo{journal}{Phys. Rev. A} \textbf{\bibinfo{volume}{63}},
  \bibinfo{pages}{062504} (\bibinfo{year}{2001}).

\bibitem[{\citenamefont{Sapirstein and Cheng}(2001)}]{sapirstein:01:pra}
\bibinfo{author}{\bibfnamefont{J.}~\bibnamefont{Sapirstein}} \bibnamefont{and}
  \bibinfo{author}{\bibfnamefont{K.~T.} \bibnamefont{Cheng}},
  \bibinfo{journal}{Phys. Rev. A} \textbf{\bibinfo{volume}{63}},
  \bibinfo{pages}{032506} (\bibinfo{year}{2001}).

\bibitem[{\citenamefont{Yerokhin and Shabaev}(2001)}]{yerokhin:01:pra}
\bibinfo{author}{\bibfnamefont{V.~A.} \bibnamefont{Yerokhin}} \bibnamefont{and}
  \bibinfo{author}{\bibfnamefont{V.~M.} \bibnamefont{Shabaev}},
  \bibinfo{journal}{Phys. Rev. A} \textbf{\bibinfo{volume}{64}},
  \bibinfo{pages}{012506} (\bibinfo{year}{2001}).

\bibitem[{\citenamefont{Sen'kov and Dmitriev}(2002)}]{sen'kov:02:npa}
\bibinfo{author}{\bibfnamefont{R.~A.} \bibnamefont{Sen'kov}} \bibnamefont{and}
  \bibinfo{author}{\bibfnamefont{V.~F.} \bibnamefont{Dmitriev}},
  \bibinfo{journal}{Nucl. Phys. A} \textbf{\bibinfo{volume}{706}},
  \bibinfo{pages}{351} (\bibinfo{year}{2002}).

\bibitem[{\citenamefont{Volotka et~al.}(2003)\citenamefont{Volotka, Shabaev,
  Plunien, and Soff}}]{volotka:03:epjd}
\bibinfo{author}{\bibfnamefont{A.~V.} \bibnamefont{Volotka}},
  \bibinfo{author}{\bibfnamefont{V.~M.} \bibnamefont{Shabaev}},
  \bibinfo{author}{\bibfnamefont{G.}~\bibnamefont{Plunien}}, \bibnamefont{and}
  \bibinfo{author}{\bibfnamefont{G.}~\bibnamefont{Soff}},
  \bibinfo{journal}{Eur. Phys. J. D} \textbf{\bibinfo{volume}{23}},
  \bibinfo{pages}{51} (\bibinfo{year}{2003}).

\bibitem[{\citenamefont{Sapirstein and Cheng}(2006)}]{sapirstein:06:pra}
\bibinfo{author}{\bibfnamefont{J.}~\bibnamefont{Sapirstein}} \bibnamefont{and}
  \bibinfo{author}{\bibfnamefont{K.~T.} \bibnamefont{Cheng}},
  \bibinfo{journal}{Phys. Rev. A} \textbf{\bibinfo{volume}{74}},
  \bibinfo{pages}{042513} (\bibinfo{year}{2006}).

\bibitem[{\citenamefont{Oreshkina et~al.}(2007)\citenamefont{Oreshkina,
  Volotka, Glazov, Tupitsyn, Shabaev, and Plunien}}]{oreshkina:07:os}
\bibinfo{author}{\bibfnamefont{N.~S.} \bibnamefont{Oreshkina}},
  \bibinfo{author}{\bibfnamefont{A.~V.} \bibnamefont{Volotka}},
  \bibinfo{author}{\bibfnamefont{D.~A.} \bibnamefont{Glazov}},
  \bibinfo{author}{\bibfnamefont{I.~I.} \bibnamefont{Tupitsyn}},
  \bibinfo{author}{\bibfnamefont{V.~M.} \bibnamefont{Shabaev}},
  \bibnamefont{and} \bibinfo{author}{\bibfnamefont{G.}~\bibnamefont{Plunien}},
  \bibinfo{journal}{Opt. Spektrosk. \textbf{102}, 889 [Opt. Spectrosc.
  \textbf{102}, 815]}  (\bibinfo{year}{2007}).

\bibitem[{\citenamefont{Kozhedub et~al.}(2007)\citenamefont{Kozhedub, Glazov,
  Artemyev, Oreshkina, Shabaev, Tupitsyn, Volotka, and
  Plunien}}]{kozhedub:07:pra}
\bibinfo{author}{\bibfnamefont{Y.~S.} \bibnamefont{Kozhedub}},
  \bibinfo{author}{\bibfnamefont{D.~A.} \bibnamefont{Glazov}},
  \bibinfo{author}{\bibfnamefont{A.~N.} \bibnamefont{Artemyev}},
  \bibinfo{author}{\bibfnamefont{N.~S.} \bibnamefont{Oreshkina}},
  \bibinfo{author}{\bibfnamefont{V.~M.} \bibnamefont{Shabaev}},
  \bibinfo{author}{\bibfnamefont{I.~I.} \bibnamefont{Tupitsyn}},
  \bibinfo{author}{\bibfnamefont{A.~V.} \bibnamefont{Volotka}},
  \bibnamefont{and} \bibinfo{author}{\bibfnamefont{G.}~\bibnamefont{Plunien}},
  \bibinfo{journal}{Phys. Rev. A} \textbf{\bibinfo{volume}{76}},
  \bibinfo{pages}{012511} (\bibinfo{year}{2007}).

\bibitem[{\citenamefont{Sapirstein and Cheng}(2008)}]{sapirstein:08:pra}
\bibinfo{author}{\bibfnamefont{J.}~\bibnamefont{Sapirstein}} \bibnamefont{and}
  \bibinfo{author}{\bibfnamefont{K.~T.} \bibnamefont{Cheng}},
  \bibinfo{journal}{Phys. Rev. A} \textbf{\bibinfo{volume}{78}},
  \bibinfo{pages}{022515} (\bibinfo{year}{2008}).

\bibitem[{\citenamefont{Volotka et~al.}(2008)\citenamefont{Volotka, Glazov,
  Tupitsyn, Oreshkina, Plunien, and Shabaev}}]{volotka:08:pra}
\bibinfo{author}{\bibfnamefont{A.~V.} \bibnamefont{Volotka}},
  \bibinfo{author}{\bibfnamefont{D.~A.} \bibnamefont{Glazov}},
  \bibinfo{author}{\bibfnamefont{I.~I.} \bibnamefont{Tupitsyn}},
  \bibinfo{author}{\bibfnamefont{N.~S.} \bibnamefont{Oreshkina}},
  \bibinfo{author}{\bibfnamefont{G.}~\bibnamefont{Plunien}}, \bibnamefont{and}
  \bibinfo{author}{\bibfnamefont{V.~M.} \bibnamefont{Shabaev}},
  \bibinfo{journal}{Phys. Rev. A} \textbf{\bibinfo{volume}{78}},
  \bibinfo{pages}{062507} (\bibinfo{year}{2008}).

\bibitem[{\citenamefont{Oreshkina et~al.}(2008)\citenamefont{Oreshkina, Glazov,
  Volotka, Shabaev, Tupitsyn, and Plunien}}]{oreshkina:08:pla}
\bibinfo{author}{\bibfnamefont{N.~S.} \bibnamefont{Oreshkina}},
  \bibinfo{author}{\bibfnamefont{D.~A.} \bibnamefont{Glazov}},
  \bibinfo{author}{\bibfnamefont{A.~V.} \bibnamefont{Volotka}},
  \bibinfo{author}{\bibfnamefont{V.~M.} \bibnamefont{Shabaev}},
  \bibinfo{author}{\bibfnamefont{I.~I.} \bibnamefont{Tupitsyn}},
  \bibnamefont{and} \bibinfo{author}{\bibfnamefont{G.}~\bibnamefont{Plunien}},
  \bibinfo{journal}{Phys. Lett. A} \textbf{\bibinfo{volume}{372}},
  \bibinfo{pages}{675} (\bibinfo{year}{2008}).

\bibitem[{\citenamefont{Volotka et~al.}(2009)\citenamefont{Volotka, Glazov,
  Shabaev, Tupitsyn, and Plunien}}]{volotka:09:prl}
\bibinfo{author}{\bibfnamefont{A.~V.} \bibnamefont{Volotka}},
  \bibinfo{author}{\bibfnamefont{D.~A.} \bibnamefont{Glazov}},
  \bibinfo{author}{\bibfnamefont{V.~M.} \bibnamefont{Shabaev}},
  \bibinfo{author}{\bibfnamefont{I.~I.} \bibnamefont{Tupitsyn}},
  \bibnamefont{and} \bibinfo{author}{\bibfnamefont{G.}~\bibnamefont{Plunien}},
  \bibinfo{journal}{Phys. Rev. Lett.} \textbf{\bibinfo{volume}{103}},
  \bibinfo{pages}{033005} (\bibinfo{year}{2009}).

\bibitem[{\citenamefont{Glazov et~al.}(2010)\citenamefont{Glazov, Volotka,
  Shabaev, Tupitsyn, and Plunien}}]{glazov:10:pra}
\bibinfo{author}{\bibfnamefont{D.~A.} \bibnamefont{Glazov}},
  \bibinfo{author}{\bibfnamefont{A.~V.} \bibnamefont{Volotka}},
  \bibinfo{author}{\bibfnamefont{V.~M.} \bibnamefont{Shabaev}},
  \bibinfo{author}{\bibfnamefont{I.~I.} \bibnamefont{Tupitsyn}},
  \bibnamefont{and} \bibinfo{author}{\bibfnamefont{G.}~\bibnamefont{Plunien}},
  \bibinfo{journal}{Phys. Rev. A} \textbf{\bibinfo{volume}{81}},
  \bibinfo{pages}{062112} (\bibinfo{year}{2010}).

\bibitem[{\citenamefont{Andreev et~al.}(2012)\citenamefont{Andreev, Glazov,
  Volotka, Shabaev, and Plunien}}]{andreev:12:pra}
\bibinfo{author}{\bibfnamefont{O.~V.} \bibnamefont{Andreev}},
  \bibinfo{author}{\bibfnamefont{D.~A.} \bibnamefont{Glazov}},
  \bibinfo{author}{\bibfnamefont{A.~V.} \bibnamefont{Volotka}},
  \bibinfo{author}{\bibfnamefont{V.~M.} \bibnamefont{Shabaev}},
  \bibnamefont{and} \bibinfo{author}{\bibfnamefont{G.}~\bibnamefont{Plunien}},
  \bibinfo{journal}{Phys. Rev. A} \textbf{\bibinfo{volume}{85}},
  \bibinfo{pages}{022510} (\bibinfo{year}{2012}).

\bibitem[{\citenamefont{Volotka et~al.}(2012)\citenamefont{Volotka, Glazov,
  Andreev, Shabaev, Tupitsyn, and Plunien}}]{volotka:12:prl}
\bibinfo{author}{\bibfnamefont{A.~V.} \bibnamefont{Volotka}},
  \bibinfo{author}{\bibfnamefont{D.~A.} \bibnamefont{Glazov}},
  \bibinfo{author}{\bibfnamefont{O.~V.} \bibnamefont{Andreev}},
  \bibinfo{author}{\bibfnamefont{V.~M.} \bibnamefont{Shabaev}},
  \bibinfo{author}{\bibfnamefont{I.~I.} \bibnamefont{Tupitsyn}},
  \bibnamefont{and} \bibinfo{author}{\bibfnamefont{G.}~\bibnamefont{Plunien}},
  \bibinfo{journal}{Phys. Rev. Lett.} \textbf{\bibinfo{volume}{108}},
  \bibinfo{pages}{073001} (\bibinfo{year}{2012}).

\bibitem[{\citenamefont{Lochmann et~al.}(2014)\citenamefont{Lochmann, J\"ohren,
  Geppert, Andelkovic, Anielski, Botermann, Bussmann, Dax, Fr\"ommgen, Hammen
  et~al.}}]{lochmann:14:pra}
\bibinfo{author}{\bibfnamefont{M.}~\bibnamefont{Lochmann}},
  \bibinfo{author}{\bibfnamefont{R.}~\bibnamefont{J\"ohren}},
  \bibinfo{author}{\bibfnamefont{C.}~\bibnamefont{Geppert}},
  \bibinfo{author}{\bibfnamefont{Z.}~\bibnamefont{Andelkovic}},
  \bibinfo{author}{\bibfnamefont{D.}~\bibnamefont{Anielski}},
  \bibinfo{author}{\bibfnamefont{B.}~\bibnamefont{Botermann}},
  \bibinfo{author}{\bibfnamefont{M.}~\bibnamefont{Bussmann}},
  \bibinfo{author}{\bibfnamefont{A.}~\bibnamefont{Dax}},
  \bibinfo{author}{\bibfnamefont{N.}~\bibnamefont{Fr\"ommgen}},
  \bibinfo{author}{\bibfnamefont{M.}~\bibnamefont{Hammen}},
  \bibnamefont{et~al.}, \bibinfo{journal}{Phys. Rev. A}
  \textbf{\bibinfo{volume}{90}}, \bibinfo{pages}{030501(R)}
  (\bibinfo{year}{2014}).

\bibitem[{\citenamefont{Ullmann et~al.}(2015)\citenamefont{Ullmann, Andelkovic,
  Dax, Geithner, Geppert, Gorges, Hammen, Hannen, Kaufmann, K\"onig
  et~al.}}]{ullmann:15:jpb}
\bibinfo{author}{\bibfnamefont{J.}~\bibnamefont{Ullmann}},
  \bibinfo{author}{\bibfnamefont{Z.}~\bibnamefont{Andelkovic}},
  \bibinfo{author}{\bibfnamefont{A.}~\bibnamefont{Dax}},
  \bibinfo{author}{\bibfnamefont{W.}~\bibnamefont{Geithner}},
  \bibinfo{author}{\bibfnamefont{C.}~\bibnamefont{Geppert}},
  \bibinfo{author}{\bibfnamefont{C.}~\bibnamefont{Gorges}},
  \bibinfo{author}{\bibfnamefont{M.}~\bibnamefont{Hammen}},
  \bibinfo{author}{\bibfnamefont{V.}~\bibnamefont{Hannen}},
  \bibinfo{author}{\bibfnamefont{S.}~\bibnamefont{Kaufmann}},
  \bibinfo{author}{\bibfnamefont{K.}~\bibnamefont{K\"onig}},
  \bibnamefont{et~al.}, \bibinfo{journal}{J. Phys. B}
  \textbf{\bibinfo{volume}{48}}, \bibinfo{pages}{144022}
  (\bibinfo{year}{2015}).

\bibitem[{\citenamefont{S\'anchez et~al.}(2017)\citenamefont{S\'anchez,
  Lochmann, J\"ohren, Andelkovic, Anielski, Botermann, Bussmann, Dax,
  Fr\"ommgen, Geppert et~al.}}]{sanchez:17:jpb}
\bibinfo{author}{\bibfnamefont{R.}~\bibnamefont{S\'anchez}},
  \bibinfo{author}{\bibfnamefont{M.}~\bibnamefont{Lochmann}},
  \bibinfo{author}{\bibfnamefont{R.}~\bibnamefont{J\"ohren}},
  \bibinfo{author}{\bibfnamefont{Z.}~\bibnamefont{Andelkovic}},
  \bibinfo{author}{\bibfnamefont{D.}~\bibnamefont{Anielski}},
  \bibinfo{author}{\bibfnamefont{B.}~\bibnamefont{Botermann}},
  \bibinfo{author}{\bibfnamefont{M.}~\bibnamefont{Bussmann}},
  \bibinfo{author}{\bibfnamefont{A.}~\bibnamefont{Dax}},
  \bibinfo{author}{\bibfnamefont{N.}~\bibnamefont{Fr\"ommgen}},
  \bibinfo{author}{\bibfnamefont{C.}~\bibnamefont{Geppert}},
  \bibnamefont{et~al.}, \bibinfo{journal}{J. Phys. B}
  \textbf{\bibinfo{volume}{50}}, \bibinfo{pages}{085004}
  (\bibinfo{year}{2017}).

\bibitem[{\citenamefont{Ullmann et~al.}(2017)\citenamefont{Ullmann, Andelkovic,
  Brandau, Dax, Geithner, Geppert, Gorges, Hammen, Hannen, Kaufmann
  et~al.}}]{ullmann:17:nc}
\bibinfo{author}{\bibfnamefont{J.}~\bibnamefont{Ullmann}},
  \bibinfo{author}{\bibfnamefont{Z.}~\bibnamefont{Andelkovic}},
  \bibinfo{author}{\bibfnamefont{C.}~\bibnamefont{Brandau}},
  \bibinfo{author}{\bibfnamefont{A.}~\bibnamefont{Dax}},
  \bibinfo{author}{\bibfnamefont{W.}~\bibnamefont{Geithner}},
  \bibinfo{author}{\bibfnamefont{C.}~\bibnamefont{Geppert}},
  \bibinfo{author}{\bibfnamefont{C.}~\bibnamefont{Gorges}},
  \bibinfo{author}{\bibfnamefont{M.}~\bibnamefont{Hammen}},
  \bibinfo{author}{\bibfnamefont{V.}~\bibnamefont{Hannen}},
  \bibinfo{author}{\bibfnamefont{S.}~\bibnamefont{Kaufmann}},
  \bibnamefont{et~al.}, \bibinfo{journal}{Nat. Commun.}
  \textbf{\bibinfo{volume}{8}}, \bibinfo{pages}{15484} (\bibinfo{year}{2017}).

\bibitem[{\citenamefont{Karr}(2017)}]{karr:17:nphys}
\bibinfo{author}{\bibfnamefont{J.~P.} \bibnamefont{Karr}},
  \bibinfo{journal}{Nat. Phys.} \textbf{\bibinfo{volume}{13}},
  \bibinfo{pages}{533} (\bibinfo{year}{2017}).

\bibitem[{\citenamefont{Skripnikov et~al.}(2018)\citenamefont{Skripnikov,
  Schmidt, Ullmann, Geppert, Kraus, Kresse, N\"ortersh\"auser, Privalov,
  Scheibe, Shabaev et~al.}}]{skripnikov:18:prl}
\bibinfo{author}{\bibfnamefont{L.~V.} \bibnamefont{Skripnikov}},
  \bibinfo{author}{\bibfnamefont{S.}~\bibnamefont{Schmidt}},
  \bibinfo{author}{\bibfnamefont{J.}~\bibnamefont{Ullmann}},
  \bibinfo{author}{\bibfnamefont{C.}~\bibnamefont{Geppert}},
  \bibinfo{author}{\bibfnamefont{F.}~\bibnamefont{Kraus}},
  \bibinfo{author}{\bibfnamefont{B.}~\bibnamefont{Kresse}},
  \bibinfo{author}{\bibfnamefont{W.}~\bibnamefont{N\"ortersh\"auser}},
  \bibinfo{author}{\bibfnamefont{A.~F.} \bibnamefont{Privalov}},
  \bibinfo{author}{\bibfnamefont{B.}~\bibnamefont{Scheibe}},
  \bibinfo{author}{\bibfnamefont{V.~M.} \bibnamefont{Shabaev}},
  \bibnamefont{et~al.}, \bibinfo{journal}{Phys. Rev. Lett.}
  \textbf{\bibinfo{volume}{120}}, \bibinfo{pages}{093001}
  (\bibinfo{year}{2018}).

\bibitem[{\citenamefont{Schmidt et~al.}(2018)\citenamefont{Schmidt, Billowes,
  Bissell, Blaum, Garcia~Ruiz, Heylen, Malbrunot-Ettenauer, Neyens,
  N\"ortersh\"auser, Plunien et~al.}}]{schmidt:18:plb}
\bibinfo{author}{\bibfnamefont{S.}~\bibnamefont{Schmidt}},
  \bibinfo{author}{\bibfnamefont{J.}~\bibnamefont{Billowes}},
  \bibinfo{author}{\bibfnamefont{M.~L.} \bibnamefont{Bissell}},
  \bibinfo{author}{\bibfnamefont{K.}~\bibnamefont{Blaum}},
  \bibinfo{author}{\bibfnamefont{R.~F.} \bibnamefont{Garcia~Ruiz}},
  \bibinfo{author}{\bibfnamefont{H.}~\bibnamefont{Heylen}},
  \bibinfo{author}{\bibfnamefont{S.}~\bibnamefont{Malbrunot-Ettenauer}},
  \bibinfo{author}{\bibfnamefont{G.}~\bibnamefont{Neyens}},
  \bibinfo{author}{\bibfnamefont{W.}~\bibnamefont{N\"ortersh\"auser}},
  \bibinfo{author}{\bibfnamefont{G.}~\bibnamefont{Plunien}},
  \bibnamefont{et~al.}, \bibinfo{journal}{Phys. Lett. B}
  \textbf{\bibinfo{volume}{779}}, \bibinfo{pages}{324} (\bibinfo{year}{2018}).

\bibitem[{\citenamefont{Verdebout et~al.}(2014)\citenamefont{Verdebout, Naz\'e,
  J\"onsson, Rynkun, Godefroid, and Gaigalas}}]{verdebout:14:adndt}
\bibinfo{author}{\bibfnamefont{S.}~\bibnamefont{Verdebout}},
  \bibinfo{author}{\bibfnamefont{C.}~\bibnamefont{Naz\'e}},
  \bibinfo{author}{\bibfnamefont{P.}~\bibnamefont{J\"onsson}},
  \bibinfo{author}{\bibfnamefont{P.}~\bibnamefont{Rynkun}},
  \bibinfo{author}{\bibfnamefont{M.}~\bibnamefont{Godefroid}},
  \bibnamefont{and} \bibinfo{author}{\bibfnamefont{G.}~\bibnamefont{Gaigalas}},
  \bibinfo{journal}{At. Data Nucl. Data Tables} \textbf{\bibinfo{volume}{100}},
  \bibinfo{pages}{1111} (\bibinfo{year}{2014}).

\bibitem[{\citenamefont{{Le~Bellac}}(1963)}]{bellac:63:np}
\bibinfo{author}{\bibfnamefont{M.}~\bibnamefont{{Le~Bellac}}},
  \bibinfo{journal}{Nucl. Phys.} \textbf{\bibinfo{volume}{40}},
  \bibinfo{pages}{645} (\bibinfo{year}{1963}).

\bibitem[{\citenamefont{Woods and Saxon}(1954)}]{woods:54:pr}
\bibinfo{author}{\bibfnamefont{R.~D.} \bibnamefont{Woods}} \bibnamefont{and}
  \bibinfo{author}{\bibfnamefont{D.~S.} \bibnamefont{Saxon}},
  \bibinfo{journal}{Phys. Rev.} \textbf{\bibinfo{volume}{95}},
  \bibinfo{pages}{577} (\bibinfo{year}{1954}).

\bibitem[{\citenamefont{Rost}(1968)}]{rost:68:plb}
\bibinfo{author}{\bibfnamefont{E.}~\bibnamefont{Rost}}, \bibinfo{journal}{Phys.
  Lett. B} \textbf{\bibinfo{volume}{26}}, \bibinfo{pages}{184}
  (\bibinfo{year}{1968}).

\bibitem[{\citenamefont{Bratzev et~al.}(1977)\citenamefont{Bratzev, Deyneka,
  and Tupitsyn}}]{bratsev:77}
\bibinfo{author}{\bibfnamefont{V.~F.} \bibnamefont{Bratzev}},
  \bibinfo{author}{\bibfnamefont{G.~B.} \bibnamefont{Deyneka}},
  \bibnamefont{and} \bibinfo{author}{\bibfnamefont{I.~I.}
  \bibnamefont{Tupitsyn}}, \bibinfo{journal}{Izv. Akad. Nauk SSSR, Ser. Fiz.
  \textbf{41}, 2655 [Bull. Acad. Sci. USSR, Phys. Ser. \textbf{41}, 173]}
  (\bibinfo{year}{1977}).

\bibitem[{\citenamefont{Tupitsyn et~al.}(2005)\citenamefont{Tupitsyn, Volotka,
  Glazov, Shabaev, Plunien, {Crespo~L\'opez-Urrutia}, Lapierre, and
  Ullrich}}]{tupitsyn:05:pra}
\bibinfo{author}{\bibfnamefont{I.~I.} \bibnamefont{Tupitsyn}},
  \bibinfo{author}{\bibfnamefont{A.~V.} \bibnamefont{Volotka}},
  \bibinfo{author}{\bibfnamefont{D.~A.} \bibnamefont{Glazov}},
  \bibinfo{author}{\bibfnamefont{V.~M.} \bibnamefont{Shabaev}},
  \bibinfo{author}{\bibfnamefont{G.}~\bibnamefont{Plunien}},
  \bibinfo{author}{\bibfnamefont{J.~R.}
  \bibnamefont{{Crespo~L\'opez-Urrutia}}},
  \bibinfo{author}{\bibfnamefont{A.}~\bibnamefont{Lapierre}}, \bibnamefont{and}
  \bibinfo{author}{\bibfnamefont{J.}~\bibnamefont{Ullrich}},
  \bibinfo{journal}{Phys. Rev. A} \textbf{\bibinfo{volume}{72}},
  \bibinfo{pages}{062503} (\bibinfo{year}{2005}).

\bibitem[{\citenamefont{Shabaev et~al.}(2004)\citenamefont{Shabaev, Tupitsyn,
  Yerokhin, Plunien, and Soff}}]{shabaev:04:prl}
\bibinfo{author}{\bibfnamefont{V.~M.} \bibnamefont{Shabaev}},
  \bibinfo{author}{\bibfnamefont{I.~I.} \bibnamefont{Tupitsyn}},
  \bibinfo{author}{\bibfnamefont{V.~A.} \bibnamefont{Yerokhin}},
  \bibinfo{author}{\bibfnamefont{G.}~\bibnamefont{Plunien}}, \bibnamefont{and}
  \bibinfo{author}{\bibfnamefont{G.}~\bibnamefont{Soff}},
  \bibinfo{journal}{Phys. Rev. Lett.} \textbf{\bibinfo{volume}{93}},
  \bibinfo{pages}{130405} (\bibinfo{year}{2004}).

\bibitem[{\citenamefont{Sapirstein and Johnson}(1996)}]{sapirstein:96:jpb}
\bibinfo{author}{\bibfnamefont{J.}~\bibnamefont{Sapirstein}} \bibnamefont{and}
  \bibinfo{author}{\bibfnamefont{W.~R.} \bibnamefont{Johnson}},
  \bibinfo{journal}{J. Phys. B} \textbf{\bibinfo{volume}{29}},
  \bibinfo{pages}{5213} (\bibinfo{year}{1996}).

\bibitem[{\citenamefont{Yerokhin and Jentschura}(2010)}]{yerokhin:10:pra}
\bibinfo{author}{\bibfnamefont{V.~A.} \bibnamefont{Yerokhin}} \bibnamefont{and}
  \bibinfo{author}{\bibfnamefont{U.~D.} \bibnamefont{Jentschura}},
  \bibinfo{journal}{Phys. Rev. A} \textbf{\bibinfo{volume}{81}},
  \bibinfo{pages}{012502} (\bibinfo{year}{2010}).

\bibitem[{\citenamefont{Stone}(2005)}]{stone:05:adndt}
\bibinfo{author}{\bibfnamefont{N.~J.} \bibnamefont{Stone}},
  \bibinfo{journal}{At. Data Nucl. Data Tables} \textbf{\bibinfo{volume}{90}},
  \bibinfo{pages}{75} (\bibinfo{year}{2005}).

\bibitem[{\citenamefont{Gustavsson and
  {M{\aa}rtensson-Pendrill}}(1998)}]{gustavsson:98:pra}
\bibinfo{author}{\bibfnamefont{M.~G.~H.} \bibnamefont{Gustavsson}}
  \bibnamefont{and} \bibinfo{author}{\bibfnamefont{A.-M.}
  \bibnamefont{{M{\aa}rtensson-Pendrill}}}, \bibinfo{journal}{Phys. Rev. A}
  \textbf{\bibinfo{volume}{58}}, \bibinfo{pages}{3611} (\bibinfo{year}{1998}).

\bibitem[{\citenamefont{Angeli and Marinova}(2013)}]{angeli:13:adndt}
\bibinfo{author}{\bibfnamefont{I.}~\bibnamefont{Angeli}} \bibnamefont{and}
  \bibinfo{author}{\bibfnamefont{K.~P.} \bibnamefont{Marinova}},
  \bibinfo{journal}{At. Data Nucl. Data Tables} \textbf{\bibinfo{volume}{99}},
  \bibinfo{pages}{69} (\bibinfo{year}{2013}).

\end{thebibliography}

%
%
%
\newpage
%
%
\begin{table}
\caption{Individual one-electron contributions to the ground-state hyperfine splitting in high-$Z$ boronlike ions.
\label{tab:A}
}
\vspace{0.5cm}
\begin{tabular}{lcllllll}
\hline
\hline
Isotope
& $I^\pi$
& \multicolumn{1}{c}{$\mu/\mu_N$}
& \multicolumn{1}{c}{$\la r^2 \ra^{1/2}$}
& \multicolumn{1}{c}{$A$}
& \multicolumn{1}{c}{$\delta$}
& \multicolumn{1}{c}{$\epssph$}
& \multicolumn{1}{c}{$\epsws$} \\
\hline
$^{113}_{49}\mathrm{In}$ & $\frac{9}{2}+$ & 5.5289(2)   & 4.6010 & 1.31236 & 0.00138(2)  & 0.00045 & 0.00041(9)   \\
$^{121}_{51}\mathrm{Sb}$ & $\frac{5}{2}+$ & 3.3634(3)   & 4.6802 & 1.34536 & 0.00169(2)  & 0.00054 & 0.00049(11)  \\
$^{123}_{51}\mathrm{Sb}$ & $\frac{7}{2}+$ & 2.5498(2)   & 4.6879 & 1.34536 & 0.00169(2)  & 0.00054 & 0.00013(11)  \\
$^{127}_{53}\mathrm{I}$  & $\frac{5}{2}+$ & 2.81327(8)  & 4.7500 & 1.38113 & 0.00206(2)  & 0.00064 & 0.00054(13)  \\
$^{133}_{55}\mathrm{Cs}$ & $\frac{7}{2}+$ & 2.58202     & 4.8041 & 1.41994 & 0.00250(2)  & 0.00076 & 0.00019(15)  \\
$^{139}_{57}\mathrm{La}$ & $\frac{7}{2}+$ & 2.78305     & 4.8550 & 1.46209 & 0.00302(3)  & 0.00089 & 0.00031(18)  \\
$^{141}_{59}\mathrm{Pr}$ & $\frac{5}{2}+$ & 4.2754(5)   & 4.8919 & 1.50793 & 0.00364(3)  & 0.00105 & 0.00097(21)  \\
$^{151}_{63}\mathrm{Eu}$ & $\frac{5}{2}+$ & 3.4717(6)   & 5.0522 & 1.61233 & 0.00533(5)  & 0.00145 & 0.00122(29)  \\
$^{159}_{65}\mathrm{Tb}$ & $\frac{3}{2}+$ & 2.014(4)    & 5.0600 & 1.67188 & 0.00637(5)  & 0.00168 & 0.00114(34)  \\
$^{165}_{67}\mathrm{Ho}$ & $\frac{7}{2}-$ & 4.177(5)    & 5.2022 & 1.73714 & 0.00777(6)  & 0.00198 & 0.00156(40)  \\
$^{175}_{71}\mathrm{Lu}$ & $\frac{7}{2}+$ & 2.2327(11)  & 5.3700 & 1.88780 & 0.01130(7)  & 0.00270 & 0.00014(54)  \\
$^{181}_{73}\mathrm{Ta}$ & $\frac{7}{2}+$ & 2.3705(7)   & 5.3507 & 1.97507 & 0.01343(9)  & 0.00310 & 0.00042(62)  \\
$^{185}_{75}\mathrm{Re}$ & $\frac{5}{2}+$ & 3.1871(3)   & 5.3596 & 2.07186 & 0.01604(10) & 0.00356 & 0.00277(71)  \\
$^{187}_{75}\mathrm{Re}$ & $\frac{5}{2}+$ & 3.2197(3)   & 5.3698 & 2.07186 & 0.01606(10) & 0.00356 & 0.00278(71)  \\
$^{203}_{81}\mathrm{Tl}$ & $\frac{1}{2}+$ & 1.62226     & 5.4666 & 2.43538 & 0.02762(15) & 0.00542 & 0.00487(108) \\
$^{205}_{81}\mathrm{Tl}$ & $\frac{1}{2}+$ & 1.63821     & 5.4759 & 2.43538 & 0.02765(15) & 0.00542 & 0.00486(108) \\
$^{207}_{82}\mathrm{Pb}$ & $\frac{1}{2}-$ & 0.592583(9) & 5.4943 & 2.50943 & 0.03028(16) & 0.00581 & 0.01014$(^{+471}_{-101})$ \\
$^{209}_{83}\mathrm{Bi}$ & $\frac{9}{2}-$ & 4.092(2)    & 5.5211 & 2.58822 & 0.03320(17) & 0.00623 & 0.00280$(^{+21}_{-112})$  \\
\hline
\hline
\end{tabular}
\end{table}
%
%
\begin{table}
\caption{One-loop QED correction to the ground-state hyperfine splitting of high-$Z$ boronlike ions in terms of the coefficient $D$, see Eq.~(\ref{eq:xD}). The results for the self-energy and vacuum-polarization contributions are given separately for the Coulomb potential and two effective screening potentials, core-Hartree (CH) and Kohn-Sham (KS).
\label{tab:QED}
}
\begin{tabular}{llllclll}
\hline
\hline
$Z$
& \multicolumn{3}{c}{$D_{\rm SE}$}
&& \multicolumn{3}{c}{$D_{\rm VP}$}
\\
\cline{2-4}
\cline{6-8}
& \multicolumn{1}{c}{Coulomb}
& \multicolumn{1}{c}{CH}
& \multicolumn{1}{c}{KS}
&& \multicolumn{1}{c}{Coulomb}
& \multicolumn{1}{c}{CH}
& \multicolumn{1}{c}{KS}
\\
\hline
49 & $-$0.1843  & $-$0.1290 & $-$0.1328 &&   0.1639 &   0.1343 &   0.1363  \\
51 & $-$0.2338  & $-$0.1704 & $-$0.1748 &&   0.1912 &   0.1578 &   0.1601  \\
53 & $-$0.2893  & $-$0.2174 & $-$0.2224 &&   0.2228 &   0.1851 &   0.1877  \\
55 & $-$0.3519  & $-$0.2708 & $-$0.2766 &&   0.2594 &   0.2169 &   0.2198  \\
57 & $-$0.4226  & $-$0.3317 & $-$0.3383 &&   0.3019 &   0.2538 &   0.2572  \\
59 & $-$0.5027  & $-$0.4011 & $-$0.4087 &&   0.3512 &   0.2967 &   0.3007  \\
63 & $-$0.6967  & $-$0.5702 & $-$0.5802 &&   0.4742 &   0.4043 &   0.4097  \\
65 & $-$0.8150  & $-$0.6736 & $-$0.6851 &&   0.5522 &   0.4726 &   0.4789  \\
67 & $-$0.9492  & $-$0.7914 & $-$0.8047 &&   0.6410 &   0.5507 &   0.5580  \\
71 & $-$1.2817  & $-$1.0837 & $-$1.1013 &&   0.8678 &   0.7504 &   0.7604  \\
73 & $-$1.4884  & $-$1.2655 & $-$1.2859 &&   1.0141 &   0.8795 &   0.8912  \\
75 & $-$1.7280  & $-$1.4766 & $-$1.5001 &&   1.1868 &   1.0322 &   1.0459  \\
81 & $-$2.7055  & $-$2.3348 & $-$2.3719 &&   1.9081 &   1.6713 &   1.6938  \\
82 & $-$2.9177  & $-$2.5142 & $-$2.5547 &&   2.0681 &   1.8132 &   1.8378  \\
83 & $-$3.1488  & $-$2.7310 & $-$2.7741 &&   2.2426 &   1.9682 &   1.9950  \\
\hline
\hline
\end{tabular}
\end{table}
%
%
\begin{table}
\caption{Individual contributions and the total values of the ground-state hyperfine splitting of high-$Z$ boronlike ions. The one-electron factor $A'=A(\aZ)(1-\delta)(1-\varepsilon)$ is obtained from the values of $A(\aZ)$, $\delta$, and $\epsws$ given in Table~\ref{tab:A}. The interelectronic-interaction contributions $B(\alpha Z)/Z$ and $C(Z,\alpha Z)/Z^2$ are evaluated according to Sec.~\ref{sec:many-el}. The QED correction $\xQED$ is found from the KS values of the self-energy and vacuum-polarization contributions (see Sec.~\ref{sec:QED} and Table~\ref{tab:QED}). The total hfs values $\Delta E$ and the reduced values $\Delta E/(\mu/\mun)$ are presented in meV. For comparison, the previously reported results are given for $^{207}_{82}$Pb and $^{209}_{83}$Bi \cite{oreshkina:08:pla}.
\label{tab:total}
}
\tabcolsep5pt
\begin{tabular}{lllllr@{}lr@{}l}
\hline
\hline
Isotope
& \multicolumn{1}{c}{$A'$}
& \multicolumn{1}{c}{$B(\alpha Z)/Z$}
& \multicolumn{1}{c}{$C(Z,\alpha Z)/Z^2$}
& \multicolumn{1}{c}{$\xQED$}
& \multicolumn{2}{c}{$\Delta E$}
& \multicolumn{2}{c}{$\Delta E/(\mu/\mun)$}
\\
\hline
$^{113}_{49}$In & 1.31001(12)  & $-$0.17206 & 0.00617(28)  &\phm 0.00001(3)  &  36&.254(10) &   6&.5571(17)  \\
$^{121}_{51}$Sb & 1.34243(15)  & $-$0.17115 & 0.00605(31)  &  $-$0.00003(3)  &  27&.634(8)  &   8&.2162(24)  \\
$^{123}_{51}$Sb & 1.34291(15)  & $-$0.17122 & 0.00606(31)  &  $-$0.00003(3)  &  19&.959(6)  &   7&.8278(23)  \\
$^{127}_{53}$I  & 1.37754(18)  & $-$0.17082 & 0.00597(35)  &  $-$0.00008(4)  &  26&.720(9)  &   9&.4978(31)  \\
$^{133}_{55}$Cs & 1.41612(21)  & $-$0.17112 & 0.00592(38)  &  $-$0.00013(4)  &  26&.923(9)  &  10&.427(4)    \\
$^{139}_{57}$La & 1.45722(26)  & $-$0.17190 & 0.00589(42)  &  $-$0.00019(5)  &  33&.340(13) &  11&.980(5)    \\
$^{141}_{59}$Pr & 1.50098(32)  & $-$0.17313 & 0.00589(47)  &  $-$0.00025(5)  &  61&.602(26) &  14&.409(6)    \\
$^{151}_{63}$Eu & 1.60178(47)  & $-$0.17752 & 0.00598(58)  &  $-$0.00040(6)  &  65&.302(34) &  18&.810(10)   \\
$^{159}_{65}$Tb & 1.65934(56)  & $-$0.18068 & 0.00607(65)  &  $-$0.00048(7)  &  47&.989(28) &  23&.828(14)   \\
$^{165}_{67}$Ho & 1.72095(69)  & $-$0.18437 & 0.00619(72)  &  $-$0.00057(7)  &  97&.077(63) &  23&.241(15)   \\
$^{175}_{71}$Lu & 1.86621(101) & $-$0.19438 & 0.00654(91)  &  $-$0.00079(8)  &  67&.170(54) &  30&.085(24)   \\
$^{181}_{73}$Ta & 1.94773(121) & $-$0.20043 & 0.00675(102) &  $-$0.00092(9)  &  81&.005(73) &  34&.172(31)   \\
$^{185}_{75}$Re & 2.03298(145) & $-$0.20689 & 0.00699(115) &  $-$0.00106(10) & 129&.59(13)  &  40&.662(41)   \\
$^{187}_{75}$Re & 2.03292(145) & $-$0.20688 & 0.00699(115) &  $-$0.00106(10) & 130&.92(13)  &  40&.661(41)   \\
$^{203}_{81}$Tl & 2.35658(256) & $-$0.23376 & 0.00801(165) &  $-$0.00158(14) & 160&.96(23)  &  99&.22(14)    \\
$^{205}_{81}$Tl & 2.35653(256) & $-$0.23375 & 0.00801(165) &  $-$0.00158(14) & 162&.54(23)  &  99&.22(14)    \\
$^{207}_{82}$Pb & 2.40877$(^{+1146}_{-246})$
                               & $-$0.23799 & 0.00815(175) &  $-$0.00167(15) &  62&.38$(^{+33}_{-9})$
                                                                                            & 105&.26$(^{+56}_{-15})$ \\
                &              &                 &            &              &  62&.23(8)$^a$
                                                                                            & 105&.01(14)$^a$         \\
$^{209}_{83}$Bi & 2.49528$(^{+53}_{-280})$
                               & $-$0.24614 & 0.00850(189) &  $-$0.00181(15) & 257&.12$(^{+22}_{-39})$
                                                                                            &  62&.83$(^{+5}_{-9})$   \\
                &              &                 &            &              & 256&.92(35)$^b$
                                                                                            &  62&.79(9)$^a$          \\
\hline
\hline
\multicolumn{9}{l}{$^a$ Ref.~\cite{oreshkina:08:pla}.}\\
\multicolumn{9}{l}{$^b$ Ref.~\cite{oreshkina:08:pla} corrected to the new value of the nuclear magnetic moment.}
\end{tabular}
\end{table}
%
%
\end{document}